\documentclass[aps,prfluids,letterpaper,10pt,groupedaddress,notitlepage,amsmath,amssymb,amsfonts]{revtex4-2}

\pdfoutput=1
\usepackage[utf8]{inputenc}
\usepackage[T1]{fontenc}

\usepackage[american]{babel}
\usepackage[autostyle=true]{csquotes}
\usepackage{graphicx}

\usepackage{xcolor}
\usepackage[normalem]{ulem}

\usepackage[colorlinks=true]{hyperref}
\usepackage[sort&compress,capitalize]{cleveref}

\begin{document}

\title{Optimal perturbations and transition energy thresholds in boundary layer shear flows}
\author{Chris Vavaliaris}
\altaffiliation{present address: Integrated Applied Math, University of New Hampshire, Durham NH 03824, USA}
\email{cv1038@wildcats.unh.edu}
\author{Miguel Beneitez} \email{beneitez@kth.se}
\author{Dan S. Henningson} \email{henning@kth.se}
\affiliation{Linn\'e FLOW Centre and Swedish e-Science Research Centre (SeRC), KTH Mechanics, Royal Institute of Technology, SE10044 Stockholm, Sweden}
\date{May 24, 2020}

\begin{abstract}
Subcritical transition to turbulence in spatially developing boundary layer flows can be triggered efficiently by finite amplitude perturbations. In this work, we employ adjoint-based optimization to identify optimal initial perturbations in the Blasius boundary layer, culminating in the computation of the subcritical transition critical energy threshold and the associated fully localized critical optimum in a spatially extended configuration, the so called minimal seed. By dynamically rescaling the variables with the local boundary layer thickness, we show that the identified edge trajectory approaches the same attracting phase space region as previously reported edge trajectories, and reaches the region more efficiently.
\end{abstract}

\maketitle

Shear flow transition to turbulence is a complex classical physics problem demonstrating rich spatiotemporal dynamics \cite{2011-avila-etal, 2016-lemoult-etal, 2016-kreilos-etal}. The inherently nonlinear process of subcritical bifurcations and the resulting formation of spatially localized states play a dominant role both in fluid dynamics \cite{2010-schneider-gibson-burke, 2018-eckhardt} and in other areas of physics \cite{2008-knobloch, 2015-knobloch}.

The study of subcritical transition in shear flows has seen significant progress by the application of concepts and methods derived from a dynamical systems viewpoint, which naturally arise from the phase space coexistence of two attracting regimes. Such concepts are the \emph{edge manifold}, which is the invariant set separating trajectories approaching the laminar and turbulent attractors, and \emph{edge states}, saddle states present on this manifold \cite{2006-skufca-yorke-eckhardt,2007-schneider-eckhardt-yorke}. Another important phase space concept is the \emph{minimal seed}, which is the point on the edge manifold that is closest to the laminar attractor in the $L^2$ (energy) norm. Perturbing the minimal seed infinitesimally along its unstable direction yields the lowest-energy initial perturbation that can trigger subcritical transition to turbulence. This framework, along with nonlinear optimization techniques
\cite{2012-foures-caulfield-schmid,2014-kerswell-pringle-willis,2018-kerswell},
has been used to find nonlinear optima and the minimal seed in spatially homogeneous flows such as
pipe flow \cite{2010-pringle-kerswell,2012-pringle-willis-kerswell},
plane Couette flow \cite{2011-monokrousos-etal,2012-rabin-caulfield-kerswell,2013-duguet-etal}
and the asymptotic suction boundary layer flow \cite{2015-cherubini-depalma-robinet}.

Previous work has shown that the application of dynamical systems concepts can be fruitful in spatially developing flows too
\cite{2010b-cherubini-etal,2011a-cherubini-etal,2011b-cherubini-etal,2012-cherubini-etal,2012-duguet-etal,2019-beneitez-etal}. The canonical example of spatially developing flows is boundary layers, and specifically the Blasius boundary layer, which approximates the incompressible high-Reynolds laminar flow over a horizontal semi-infinite flat plate \cite{2001-schmid-henningson, 2017-schlichting-gersten}. No fixed length scale exists and the Reynolds number increases along the streamwise direction $x$; it is usually defined with the boundary layer's displacement thickness
$\delta^*(x) := \int_0^\infty (1-U(x,y)/U_\infty)dy$,
where $U$ denotes the streamwise velocity field, $U_\infty$ the freestream velocity, and $y$ the wall-normal direction. The Blasius solution is unstable to perturbations of sufficiently high amplitude, but becomes linearly unstable too at $R_{\delta^*}\simeq 520$ \cite{1970-jordinson, 1998-berlin}. Thus, depending on the local Reynolds number, transition can be triggered both subcritically and supercritically. A key difference between the two transition routes is their characteristic timescales, with the subcritical route being faster \cite{2019-beneitez-etal}. Exploiting this timescale separation, we focus on Blasius transition for low-to-moderate $R_{\delta^*}$ values, where the subcritical route is dominant.

Previous work has attempted to identify optimal initial perturbations for the Blasius boundary layer \cite{2011b-cherubini-etal}. Although these authors do indeed find nonlinear nonlocalized optima in a narrow domain, they employ a different definition for the minimal seed. More specifically, they identify a \textquote{basic building block} \cite{2011b-cherubini-etal} at the core of the found optima for initial energies above a \textquote{nonlinearity threshold} \cite{2011b-cherubini-etal}. This, however, is not the optimal initial perturbation at the critical energy threshold for transition, and is thus not consistent with the above definition.

In this work, we identify the critical energy threshold and fully localized minimal seed for subcritical transition in a spatially developing boundary layer flow. Moreover, we investigate the edge manifold's dynamics right from its point closest to the laminar solution, and demonstrate that trajectories starting from different regions of the manifold do indeed approach the same attracting region in phase space.

Within the subcritical transition timescales, we use adjoint based optimization to identify optimal initial perturbations and the critical energy threshold $E_c$ for transition. We employ a spatially extended domain, allowing for full localization of the perturbations. We timemarch the Blasius solution until a converged state is reached and use it as the baseflow; we consider the evolution of perturbations about this baseflow. For the optimization we employ a well established variational framework \cite{2012-foures-caulfield-schmid,2014-kerswell-pringle-willis,2018-kerswell}. The optimizing functional is the perturbation kinetic energy gain at time horizon $T$
\begin{equation*}
    G_T := E_T/E_0 = \left\|\mathbf{u}\left(T,\mathbf{x}\right)\right\|^2 \,/\, \left\|\mathbf{u}\left(0,\mathbf{x}\right)\right\|^2 \,,
\end{equation*}
where $\mathbf{u}(t,\mathbf{x})$ denotes the perturbation velocity field and $\|\cdot\|$ the $L^2$ norm. The governing equations are the Navier-Stokes and continuity equations. To update the initial field $\mathbf{u}_0$ we employ the gradient-rotation method \cite{2013-foures-caulfield-schmid}, which ensures that $E_0$ remains constant without an explicit energy constraint, while to relax the high memory demand during the adjoint integration we incorporate a memory checkpointing routine. The simulations are done using the spectral element code Nek5000 \cite{nek5000} and specifically the fully nonlinear adjoint based optimization codebase developed in \cite{2019-rinaldi-canton-schlatter}, which marked the first nonlinear optimization done with Nek5000 implementing the gradient rotation and memory checkpointing routines. The use of state-of-the-art methods is critical in such a computationally demanding undertaking: compared to previous studies \cite{2011b-cherubini-etal}, we use a 2.5-times longer and 10-times wider domain, as well as a 5-times longer optimization time horizon. Additionally, on average $154$ optimization iterations are performed for each reported optimum. Numerical convergence is checked by the residual
$r_n :=  \left\|(\delta_{\mathbf{u}_0}\mathcal{L}_n)^\perp\right\|^2 / \left\|\delta_{\mathbf{u}_0}\mathcal{L}_n\right\|^2$,
where $\mathcal{L}$ is the problem's Lagrangian \cite{2013-foures-caulfield-schmid}.

The parameters explicitly appearing in the algorithm are the initial energy $E_0$, the time horizon $T$, and the inflow Reynolds number $R_{in}$. Due to the spatial variation, the streamwise domain length $L_x$ implicitly affects the optimization as well. For fixed $R_{in}$, the values of $T$ and $L_x$ control the range of local Reynolds numbers the flow evolution will span. Therefore, they determine the dynamical behavior present within the optimization horizon.

\begin{figure}[tb]
    \begin{tabular}{c c}
        \includegraphics[width=.49\columnwidth]{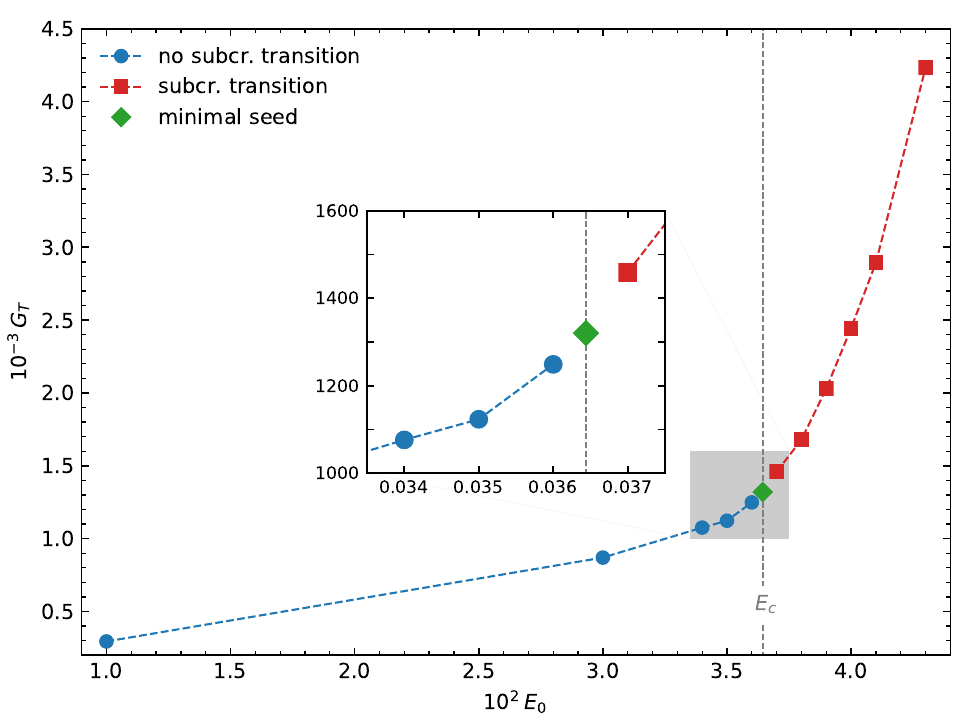} &
        \includegraphics[width=.49\columnwidth]{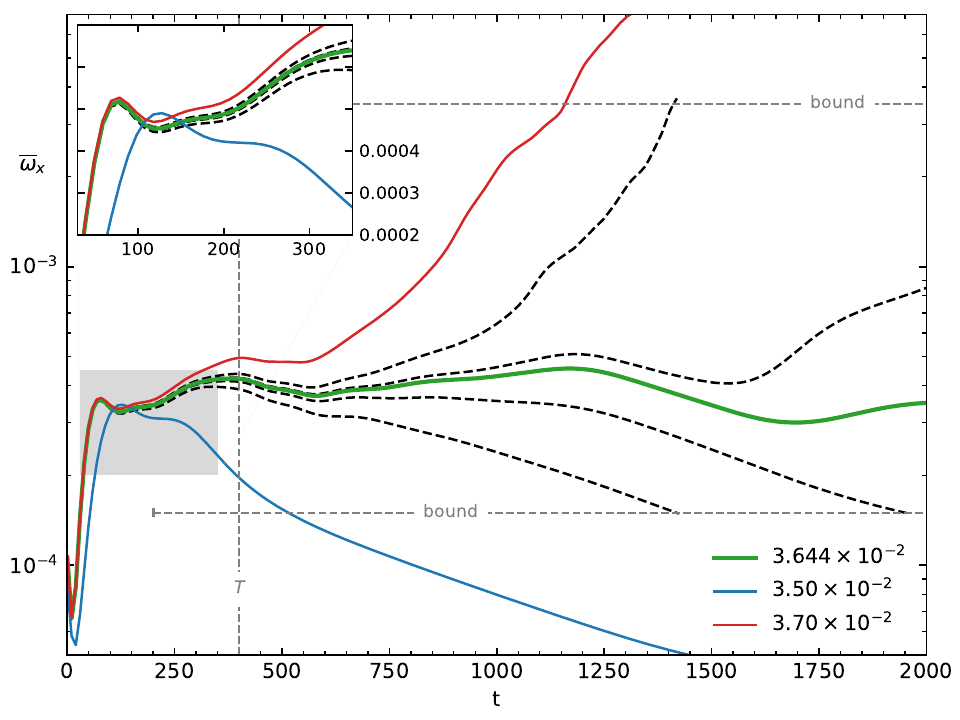} \\
        (a) & (b)
    \end{tabular}
    \caption{(a) Energy-gain diagram. The points mark the optima found for each $E_0$ level; red squares trigger subcritical transition, blue circles do not. The critical energy threshold $E_c$ and minimal seed are given by the green diamond. The shaded area is shown in the inset. (b) Edge tracking results. The green line is the identified edge trajectory; the blue and red lines correspond to optima below and above $E_c$ (see legend). The laminar and turbulent vorticity bounds used are given by the horizontal gray dashed lines; their enforcement begins at $t=200$. Their values are based on those successfully used in \cite{2019-beneitez-etal} and are determined in a trial and error process, to ensure that perturbations have sufficient time to clearly evolve towards transition or relaminarization, but to also prevent unnecessarily long simulation times. The black dashed lines are some of the trajectories ruled out during the edge tracking. The shaded area is shown in the inset.}
    \label{fig:optimization}
\end{figure}

The streamwise, normal and spanwise directions are denoted by $x$, $y$ and $z$, respectively. The presented results are obtained for $R_{in}=275$, $T=400$ and domain dimensions $(L_x, L_y, L_z)=(500, 60, 100)$. An inflow fringe region of length $l_x=25$ is used, with $l_x$ being part of $L_x$. The length and velocity scales used for nondimensionalization are the inflow boundary layer displacement thickness $\delta^*_0$ (upstream of the fringe) and $U_\infty$; the used timescale is thus $\delta^*_0/U_\infty$. The reported $R_{in}$ value is attained right downstream of the fringe. In terms of boundary conditions, at the inflow we set the velocity to its base value (zero perturbation), while a no-slip condition is imposed at the bottom wall. A stress-free (Neumann) condition is imposed at the outflow. At the top boundary, the streamwise and spanwise velocities are set to their base values (zero perturbation), and a stress free condition is used for the normal velocity. Periodicity is imposed in the spanwise direction. The derivation of the corresponding boundary conditions for the adjoint equation is presented in full in \cite{2017-rahkola}. The initial perturbation field consists of a divergence-free pair of tilted streamwise vortices, so that no symmetries are introduced into the resulting solutions \cite{1993-henningson-lundbladh-johansson}.

We do not require transition to be triggered within the time horizon $T$. To establish whether or not an optimum triggers transition, we perform long-time direct runs with time horizon $t_f=1500$ ($L_x=2500$; $L_y$, $L_z$ as above), using the volume-averaged streamwise vorticity
$\overline{\omega}_x := (1/V\int|\omega_x|^2dv)^{1/2}$
as an observable, where $dv=dxdydz$ and $V$ is the domain's volume. $\overline{\omega}_x(t)$ characterizes the streamwise vortices' strength, and is equal to zero for the baseflow. Transitioning optima present an ever increasing $\overline{\omega}_x(t)$ evolution; nontransitioning ones transiently grow and then decay while slowly relaminarizing.

To calculate $E_c$ we follow a bisection process. Starting from two $E_0$ levels, one leading to subcritical transition $E_{0,T}$ and one to relaminarization $E_{0,L}$, we choose their average value, $E_{0,i}=1/2(E_{0,T}+E_{0,L})$, and perform an optimization for this energy level. If the calculated optimum triggers transition, then $E_{0,T}\leftarrow E_{0,i}$; otherwise $E_{0,L}\leftarrow E_{0,i}$. We then set $E_{0,i+1}$ in the same fashion and continue the iterative process
\footnote{This implicitly assumes that the found optimum completely characterizes its energy level $E_0$, meaning that the \emph{global} optimum is found. Due to nonconvexity this is not guaranteed, so various computational measures have been taken to test whether this is true, and to probe the robustness of the reported results: on each level $E_0$, (1) we have reinitialized the algorithm with suboptimal solutions, as well as with optimal and suboptimal solutions of different energy levels after rescaling them; (2) we have forced the algorithm to consider suboptimal directions, to see whether the same optimum is eventually recovered. None of these measures led to solutions with higher gains than the ones reported here.}.
Having approached $E_c$ with an accuracy of $\mathcal{O}(10^{-3})$, we employ an edge tracking algorithm \cite{2001-itano-toh,2006-skufca-yorke-eckhardt,2019-beneitez-etal} to refine it and ensure that the calculated critical threshold and optimum do indeed lie on the edge manifold. The edge tracking is performed using the in-house collocation code SIMSON \cite{2007-simson}. During the edge tracking no perturbation shape optimization takes place; the only parameter changing is $E_0$ for fixed shape. The resulting $E_c$ is thus an approximation of the exact value that can only be computed by resuming the optimization runs on different $E_0$ levels all the way to $E_c$ (see e.g. \cite{2012-rabin-caulfield-kerswell}), which is currently intractable due to the associated computational cost.

\cref{fig:optimization}a presents the optimization results, with each point corresponding to the identified optimal initial perturbation. Increasing $E_0$ leads to increased gain, although with different increasing rates for energies below and above $E_c$. For $E_0<E_c$ the gain increases slowly with $E_0$, with the rate getting larger as $E_c$ is approached. For $E_0>E_c$ the gain increases rapidly with $E_0$, signifying that a qualitatively different range of dynamics is now within reach. Crossing the energy threshold $E_c$ means that the optima can now access the turbulent dynamics, leading to the attainment of highly energetic endflows via a much more effective extraction of energy from the baseflow.

We identify the critical energy threshold separating the optima undergoing subcritical transition and those that do not, converging to the value $E_c=3.644\cdot 10^{-2} \pm 2\cdot 10^{-5}$. $E_c$ is obtained by an edge tracking computation, resulting in the closest edge-bracketing trajectories presenting a separation of $2\%$ at $t=824.6$
\footnote{The $E_c$ error bounds ($\pm 2\cdot 10^{-5}$) are governed by the edge tracking domain's streamwise length $L_x$, due to the spatially developing nature of the flow.}.
In phase space terms, $E_c$ characterizes the lowest-energy iso-hypersurface that intersects the edge manifold; the minimal seed is the initial perturbation corresponding to the intersection point. The minimal seed identified in the present configuration is the point on the edge manifold that is closest to the laminar solution at $R_{\delta^*}(x_m)=425.1$, where $x_m$ corresponds to the center of mass of the optimum's spanwise velocity. Due to the spatial development, $E_c$ and the minimal seed depend on the local streamwise coordinate $x$, since $x$ sets the local Reynolds number $R_{\delta^*}(x)$.

\begin{figure}[tb]
    \begin{tabular}{c c c}
        \includegraphics[width=.28\columnwidth]{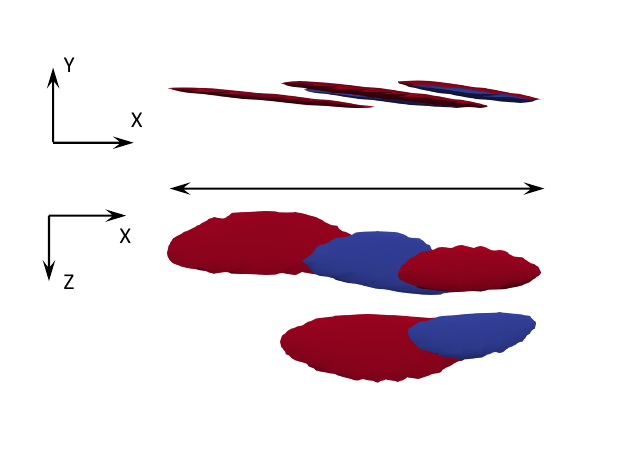} &
        \includegraphics[width=.31\columnwidth]{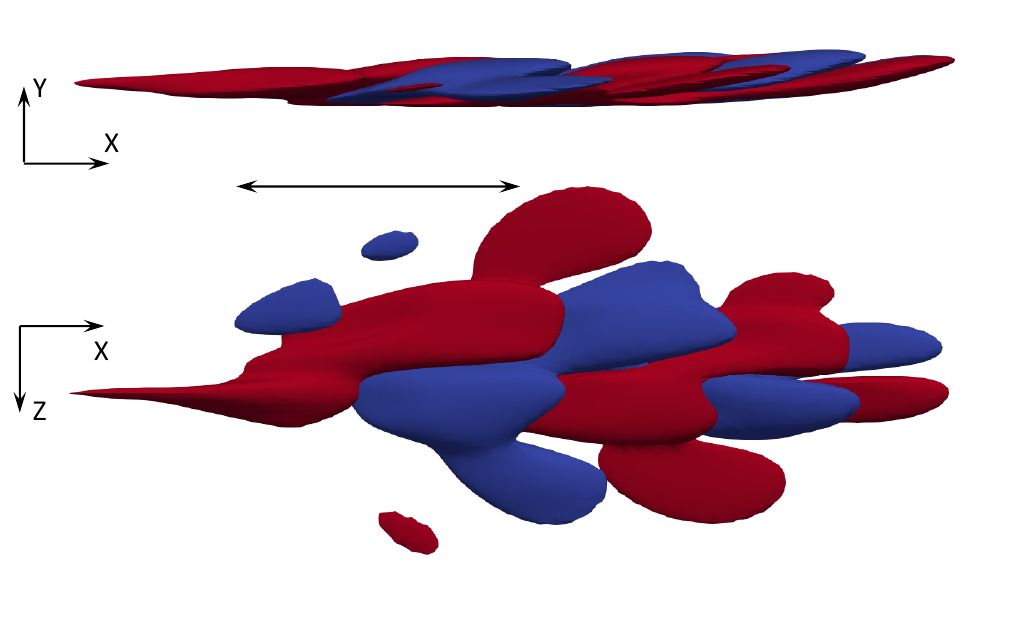} &
        \includegraphics[width=.36\columnwidth]{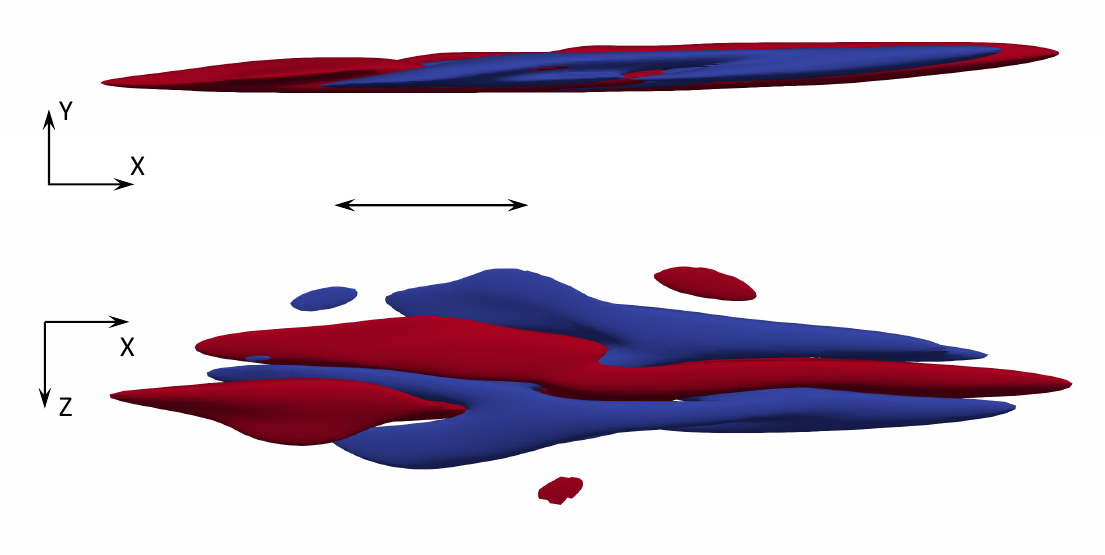} \\
        (a) & (b) & (c)
    \end{tabular}
    \begin{tabular}{c}
        \includegraphics[width=.95\columnwidth]{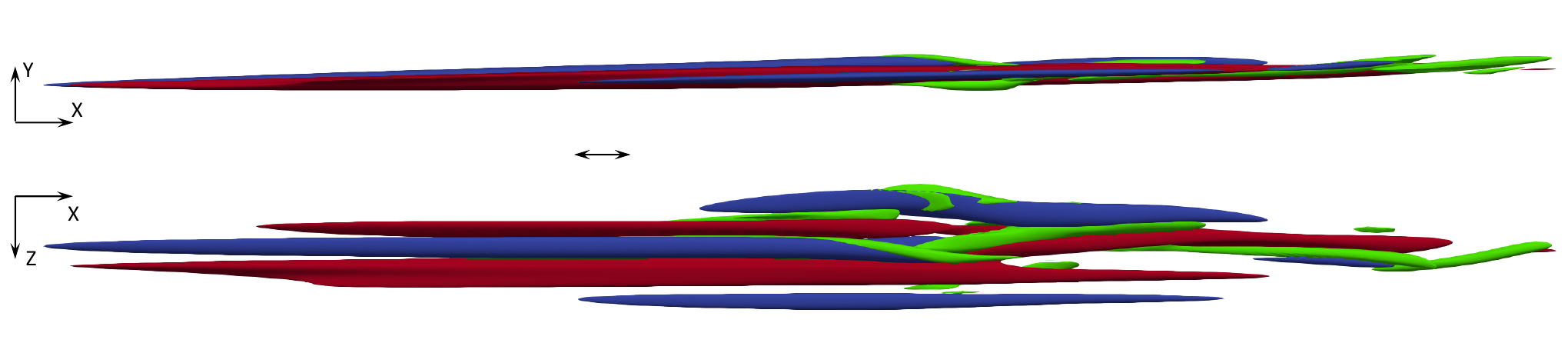} \\
        (d)
    \end{tabular}
    \caption{Streamwise perturbation velocity (blue for negative; red for positive) and $\lambda_2$ vortex criterion (green) contours of the minimal seed's evolution. Side view in top half, top view in bottom half; flow towards the right. Spatial dimensions between different subfigures not to scale. The double arrow gives the minimal seed's streamwise length ($22$ units). Contour levels for (a)-(c) $u=-8\cdot 10^{-3}$, $6\cdot 10^{-3}$; (d) $u=-6\cdot 10^{-2}$, $5\cdot 10^{-2}$, $\lambda_2=-1\cdot 10^{-5}$. (a) $t=0$; the minimal seed is nonsymmetric and backwards leaning. (b) $t=60$; the perturbation is leaning forward after the Orr mechanism's effect. (c) $t=150$; the perturbation is getting elongated and consists of high- and low-speed streaks created by the liftup mechanism. (d) $t=2500$; the perturbation is considerably longer, while maintaining its streaky structure; the $\lambda_2$ contours help identify its active core \cite{2012-duguet-etal}, typical of edge trajectory perturbations.}
    \label{fig:snapshots}
\end{figure}

\cref{fig:optimization}b shows the edge tracking results, with the green solid line being the identified edge trajectory; the vorticity curves for the optima for $E_0=3.5\cdot 10^{-2}$, $3.7\cdot 10^{-2}$ are also shown. Despite their small separation in $G_T$, the optima quickly diverge from one another and from the edge trajectory in their vorticity timeseries. Moreover, their early-time vorticity peaks are reached at different times, a result of their different asymptotic behavior and spatial positioning. Since for $E_0<E_c$ triggering subcritical transition is not possible, the corresponding optimum tries to delay its inevitable viscous decay to achieve the maximum $G_T$ possible. To do so, it also positions itself further downstream, where the local $R_{\delta^*}(x)$ values are higher, leading to more favorable growth conditions. On the other hand, transition can be triggered for $E_0>E_c$, thus the respective optimal perturbation tries to trigger it as early as possible, to take advantage of the turbulent dynamics and attain a higher $G_T$. Due to its higher energy content and its ability to trigger transition, it can also survive relatively lower $R_{\delta^*}(x)$ values; thus, it positions itself further upstream, leaving the bigger part of the domain for its energy growth and spatial spreading. Because of the spatial variation, the streamwise position the localized optima will assume within the domain is also an output of the optimization process, and is not \emph{a priori} set by the algorithm.

It is worthwhile comparing \cref{fig:optimization}a to the energy-gain diagram found for plane Couette flow (Fig. 3a in \cite{2012-rabin-caulfield-kerswell}). First, the gain discontinuity at $E_c$ found in \cite{2012-rabin-caulfield-kerswell} does not appear in the present study. This occurs because we do not require transition to be triggered within the optimization time horizon, so no turbulent endflow needs to be attained. Thus, optima on either side of the edge manifold do not diverge considerably from it, and hence are not so strongly separated in terms of gain at $t=T$. Second, $G_T$ keeps increasing for $E_0>E_c$. In \cite{2012-rabin-caulfield-kerswell}, the final energy $E_T$ saturates for $E_0\geq E_c$, leading to the decrease of gain for higher $E_0$. On the contrary, in our case $E_T$ does not saturate, because no turbulent domain-filling endflow is attained; the higher the $E_0$, the higher the resulting $E_T$. The gain keeps increasing for $E_0\gtrsim E_c$ because $E_T$ grows \textquote{more quickly} than $E_0$ as the latter is increased.

The minimal seed is fully localized and consists of alternating areas of positive and negative streamwise velocity tilted against the shear (\cref{fig:snapshots}a). Additionally, it is not spanwise symmetric, which is the only symmetric direction of the underlying baseflow, due to the nonsymmetric perturbation used to initiate the optimization algorithm. If the introduced initial perturbation is also spanwise symmetric, then the spanwise symmetry cannot be broken and the simulation is confined to a spanwise symmetric subset of solutions, which is the case in \cite{2012-duguet-etal}. Due to the baseflow's spanwise symmetry, the spanwise mirrored minimal seed satisfies all boundary conditions, and is expected to be an equivalent solution to the same optimization problem.

\begin{figure}[tb]
    \begin{tabular}{c c}
        \includegraphics[width=.51\columnwidth]{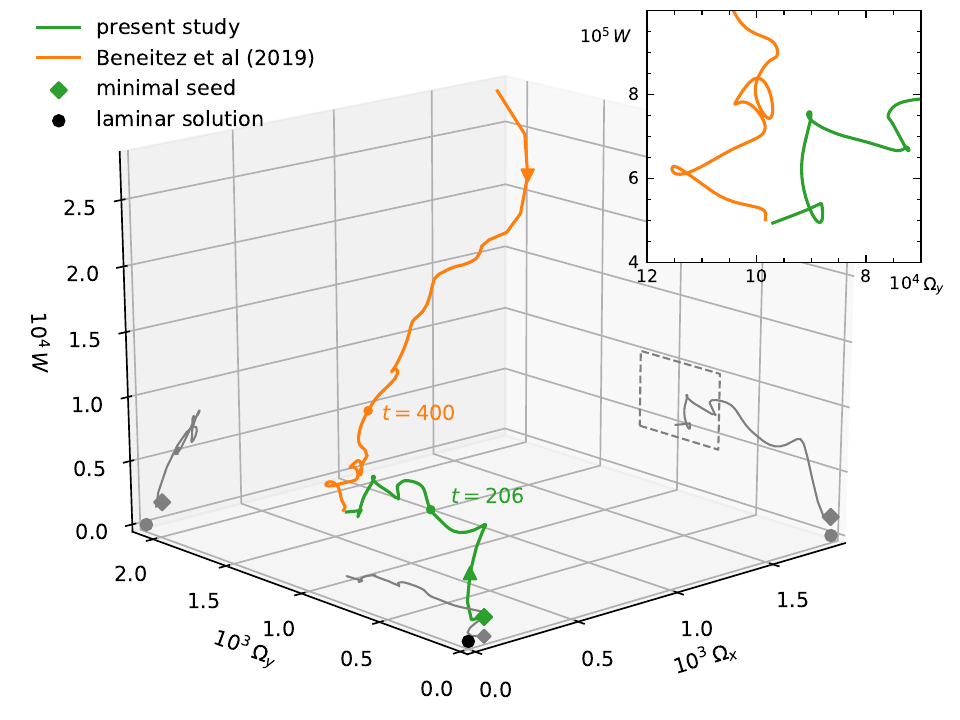} &
        \includegraphics[width=.47\columnwidth]{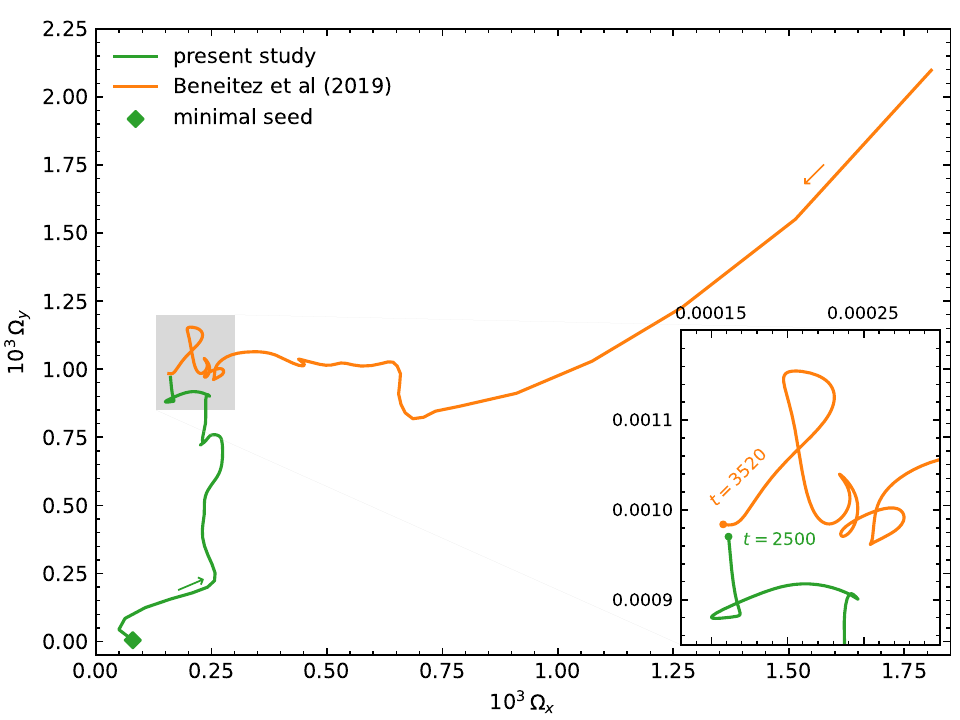} \\
        (a) & (b)
    \end{tabular}
    \caption{Arrows indicate the time direction. (a) Phase portrait of the dynamically rescaled edge trajectories identified in the present study (green line in \cref{fig:optimization}b) and in \cite{2019-beneitez-etal}. 2D projections of the green trajectory are given in gray; the area marked by the dashed rectangle is shown in the inset. (b) 2D projection of the phase portrait onto the $(\Omega_x, \Omega_y)$ plane; laminar fixed point located at the origin. The shaded area is shown in the inset.}
    \label{fig:phase-space}
\end{figure}

Investigating the minimal seed's evolution, we identify the occurrence of the Orr \cite{1907b-orr} and liftup \cite{1980-landahl} energy growth mechanisms. The Orr is a linear mechanism responsible for the early energy growth experienced by the perturbation, $t\in[0,60]$, and can be characterized by the tilting of the structure about the spanwise axis. \cref{fig:snapshots}a-b demonstrate that, although initially leaning backwards, the perturbation has been rotated and is leaning forward at $t=60$. The Orr is followed by the liftup mechanism, which acts via streamwise vortices displacing areas of high and low streamwise-momentum fluid in the wall-normal direction. This creates distinct areas of low- and high-speed fluid, leading to the formation of streaks (spanwise variation of streamwise velocity) and overall elongation of the advected structure \cite{2001-matsubara-alfredsson}; the high- and low-speed streaks making up the perturbation can be observed in \cref{fig:snapshots}c-d. Both the tilting of the perturbation caused by the Orr mechanism for $t\in[0,60]$ and the elongation caused by the liftup mechanism for $t\in[70,200]$ can be seen in the short-time simulation movies available in the supplement
\footnote{See supplemental material available
\href{https://journals.aps.org/prfluids/supplemental/10.1103/PhysRevFluids.5.062401}{online}
for two short time movies ($xy$ and $xz$ planes), $t\in[0,200]$, and one long time movie ($xz$ plane), $t\in[450,2500]$, created using the minimal seed's simulation results. The contour levels in the short time movies are as in \cref{fig:snapshots}a-c; for the long time one, velocity is as in \cref{fig:snapshots}d and $\lambda_2=-2.5\cdot 10^{-5}$. In all three movies the camera is moving with constant speed, allowing one to observe the changes in advection speed experienced by the perturbation.}.
The occurrence of the Orr and liftup mechanisms during the early energy growth stages of optima has also been reported in the past for plane Couette \cite{2013-duguet-etal}, asymptotic suction boundary layer \cite{2015-cherubini-depalma-robinet}, and Blasius boundary layer \cite{2011b-cherubini-etal,2012-cherubini-etal} flows. For longer times, we observe an approximately self-similar evolution, with the perturbation maintaining its streaky structure while growing considerably in size due to the boundary layer's spatial growth, and resembling the edge trajectory identified in \cite{2019-beneitez-etal} (\cref{fig:snapshots}d).

To investigate the edge manifold's phase space dynamics we use the volume-averaged streamwise vorticity $\overline{\omega}_x$, normal vorticity
$\overline{\omega}_y := (1/V\int|\omega_y|^2 dv)^{1/2}$,
and spanwise velocity
$\overline{w} := (1/V\int|w|^2 dv)^{1/2}$.
$\overline{\omega}_y$ is a measure of the streaks' amplitude, while $\overline{w}$ measures the lateral activity of the perturbation; all three quantities are zero for the laminar baseflow. In order to account for the boundary layer's spatial growth, we dynamically rescale the independent variables with the local displacement thickness $\delta^*(x)$ and use the rescaled phase variables $(i=x,y)$
$\Omega_i := \left(\delta^*_0/\delta^*(x)\right)^{1/2} \, \overline{\omega_i}$,
$W := \left(\delta^*_0/\delta^*(x)\right)^{3/2} \, \overline{w} \,$
\cite{2012-duguet-etal, 2019-beneitez-etal}. To get the local values $\delta^*(x(t))$ for the optimum's evolution, we calculate its spanwise velocity's center of mass $x_m(t)$, and derive an $11$-order least squares polynomial fitting
\footnote{We compute $x_m(t)$ every $1$ time unit for $t\in[0,850]$, and every $5$ time units for $t\in[855,2500]$.}.
Using the fitted polynomial $p_m(t)$, we calculate the values $\delta^*(p_m(t))\simeq\delta^*(x_m(t))$.

The rescaled edge trajectory is shown in \cref{fig:phase-space}, along with part of the edge trajectory found in \cite{2019-beneitez-etal} using a nonoptimized initial perturbation. Both trajectories approach the same relative attractor in phase space, despite their different initial conditions and early-time behavior. It has been shown in \cite{2019-beneitez-etal} that the corresponding edge trajectory stays in the identified attracting region for times up to $t\simeq 8000$. Moreover, both trajectories form similar loops in the same phase space region and mirror one another in the last part of their dynamics.  These results indicate that memory of the trajectories' initial conditions has indeed been lost \cite{2012-duguet-etal}. The formation of loops was also identified in \cite{2012-duguet-etal, 2019-beneitez-etal}, and stems from a self-sustained physical process alternating between the generation of vorticity and the creation of low-speed streaks \cite{2012-duguet-etal}. The stages of this process can also be identified by the phases of acceleration and deceleration the advected perturbation goes through, captured in the long-time simulation movie available online. The timestamps provided in \cref{fig:phase-space}a show that the minimal seed needs about half the time to reach the attracting region, compared to the nonoptimized initial perturbation; it also requires an approximately $1700$-times lower initial energy level, demonstrating its effectiveness in accessing the edge manifold's dynamics.

To summarize, we have used adjoint based optimization together with edge tracking to calculate optimal initial perturbations for subcritical transition in a spatially developing boundary layer flow, and have estimated the critical energy threshold $E_c=3.644\cdot 10^{-2} \pm 2\cdot 10^{-5}$ corresponding to the minimal seed at $R_{\delta^*}(x_m)=425.1$. We have observed the main mechanisms driving the minimal seed's early energy growth, and explained the differences in the positioning between localized optima found above and below $E_c$ in terms of the optimization process. Finally, by dynamically rescaling the variables, we have shown that two very different initial conditions on the edge manifold approach the same attracting phase space region already identified for longer times \cite{2019-beneitez-etal}. The minimal seed reaches this region more efficiently than nonoptimized initial conditions in terms of both initial energy and time needed. Future work includes computing $E_c$ for additional local Reynolds numbers $R_{\delta^*}(x)$ to reveal the scaling relationship of the two, as has already been done for parallel flow configurations \cite{2013-duguet-etal,2015-cherubini-depalma-robinet}.

\begin{acknowledgments}
We thank Yohann Duguet for helpful discussions, and Enrico Rinaldi, Jacopo Canton and Philipp Schlatter for developing the used numerical codebase. Financial support by the Swedish Research Council (VR) grant no. 2016-03541 and computing time provided by the Swedish National Infrastructure for Computing (SNIC) are gratefully acknowledged. The open source projects
\href{https://numpy.org/}{NumPy},
\href{https://www.scipy.org/}{SciPy},
\href{https://matplotlib.org/}{Matplotlib},
\href{https://ffmpeg.org/}{FFmpeg}
and
\href{https://www.paraview.org/}{ParaView}
have been used for this work.
\end{acknowledgments}

\bibliography{main}

\end{document}